\documentclass[10pt,aps,twocolumn,prd,showpacs,nofootinbib]{revtex4-1}
\usepackage{amsmath} 
\usepackage{amssymb}

\begin{document}

\title{Lorentz violation in the gravity sector: The $t$ puzzle}

\author{Yuri Bonder}
\email{bonder@nucleares.unam.mx}
\affiliation{Physics Department, Indiana University, Bloomington, IN 47405, USA}
\affiliation{Instituto de Ciencias Nucleares, Universidad Nacional Aut\'onoma de M\'exico\\
Apartado Postal 70-543, C.P. 04510, D.F., M\'exico}

\begin{abstract}
Lorentz violation is a candidate quantum-gravity signal, and the Standard-Model Extension (SME) is a widely used parametrization of such a violation. In the gravitational SME sector, there is an elusive coefficient for which no effects have been found. This is known as the $t$ puzzle and, to date, it has no compelling explanation. This paper analyzes whether there is a fundamental explanation for the $t$ puzzle. To tackle this question, several approaches are followed. Mainly, redefinitions of the dynamical fields are studied, showing that other SME coefficients can be moved to nongravitational sectors. It is also found that the gravity SME sector can be consistently treated \textit{\`a la} Palatini, and that, in the presence of spacetime boundaries, it is possible to correct its action to get the desired equations of motion. Moreover, through a reformulation as a Lanczos-type tensor, some problematic features of the $t$ term, which should arise at the phenomenological level, are revealed. The most important conclusion of the paper is that there is no evidence of a fundamental explanation for the $t$ puzzle, suggesting that it may be linked to the approximations taken at the phenomenological level.
\end{abstract}

\pacs{11.30.Cp,04.80.Cc}

\maketitle

\section{Introduction}

General relativity is the simplest gravity theory that incorporates Einstein's guiding principles, has a consistent mathematical structure, and, with the inclusion of dark matter and a cosmological constant, successfully accounts for all experimental tests \cite{Will}. However, it does not seem to be adaptable to the quantum realm, suggesting that it should be replaced by a more fundamental theory \cite{Ashtekar}. A viable strategy towards a quantum-compatible gravity theory, and which encompasses this work, is to test general relativity, through its underlying principles, with the hope of uncovering evidence of where the theory fails.

Local Lorentz invariance is one of the fundamental principles of general relativity, and the search for its violations---Lorentz violation, for short---has spawned an important amount of work. In addition, several mechanisms to generate Lorentz violation have been put forward by several quantum-gravity candidates, like string theory \cite{LV strings}, loop quantum gravity \cite{LV LQG}, and noncommutative geometries \cite{LV noncommutativity}. Moreover, it has also been suggested that Lorentz violation could arise through nonminimal gravitational couplings \cite{QGP}, by explicit symmetry-breaking mechanisms \cite{LV SSB}, and as the effects associated with generalized geometrical structures \cite{Finsler}.

To systematically test the validity of a physical principle, it is useful to have a parametrization of its possible violations. In the case of Lorentz violation, the general parametrization, called the Standard-Model Extension (SME), was originally proposed, in the context of flat spacetime, by Colladay and Kosteleck\'y \cite{ColladayKostelecky}. Since then, the SME has motivated many experiments, covering a wide range of systems, none of which has found convincing evidence for Lorentz violation. Still, these experiments have been used to set bounds on SME parameters, which are collected in Ref.~\onlinecite{DataTables}. 

The SME is conceived as an effective field theory \cite{KosteleckyPotting}, therefore, it contains all known physics plus additional Lorentz-violating terms. These terms consist of a Lorentz-violating operator, built with conventional fields, and a controlling coefficient, called an SME coefficient. In addition, since the SME contains Lorentz-violating extensions to all known physics, it can be naturally divided into sectors.

The focus of this paper is the gravitational SME sector, with vanishing torsion and cosmological constant, and where the relevant field is the spacetime metric $g_{ab}$. In particular, attention is restricted to the minimal gravity SME sector (mgSME) where the SME coefficients are directly contracted with the Riemann curvature tensor ${R_{abc}}^d$. Concretely, the mgSME action takes the form
 \begin{equation}
 S_{\rm mgSME} = S_{\rm EH} + S_{\rm LV} + S_{\rm coef} + S_{\rm matter} , \label{initial action}
 \end{equation}
with
\begin{eqnarray}
 S_{\rm EH}(g) &=& \frac{1}{2\kappa}\int d^4 x \sqrt{-g} R, \\
 S_{\rm LV}(g,k) &=& \frac{1}{2\kappa}\int d^4 x \sqrt{-g} k^{abcd} R_{abcd},
\end{eqnarray}
where $k^{abcd}$ are the corresponding SME coefficients and $\kappa$ is the standard coupling constant of general relativity. The possibility of having additional matter fields, generically denoted by $\phi$, is included through the action term $S_{\rm matter}= S_{\rm matter}(g,\phi)$. Moreover, to avoid inconsistencies with the Bianchi identity, in curved spacetimes, Lorentz violation must arise spontaneously \cite{Kostelecky04}, and the dynamics of the SME coefficients is determined by $S_{\rm coef}=S_{\rm coef}(g,k)$, which has produced some intriguing results concerning the corresponding Nambu-Goldstone modes \cite{NG}.

The Riemann tensor can be decomposed (irreducibly) by the Weyl tensor, $W_{abcd}$, the traceless Ricci tensor, $R_{ab}^T = R_{ab} - g_{ab}R/4$, and the curvature scalar, $R$. This decomposition induces a decomposition of $k^{abcd}$ into $u$, $s^{ab}$, and $t^{abcd}$, where $s^{ab}$ is symmetric and traceless and $t^{abcd}$ shares all the index symmetries of the Riemann tensor and is traceless. This allows us to verify that the number of independent components of $t^{abcd}$, $s^{ab}$, and $u$ is, respectively, $10$, $9$, and $1$. This number of components adds up to $20$, which coincides with the number of independent components of $k^{abcd}$. The explicit relation between $k^{abcd}$ and its irreducible pieces is
\begin{equation}
k^{abcd}= t^{abcd} + \frac{1}{2}\left(g^{a[c} s^{d]b}-g^{b[c} s^{d]a} \right) - g^{a[c} g^{d]b}u.
\end{equation}
Using the Riemann decomposition, it is possible to verify that
\begin{eqnarray}
k^{abcd} R_{abcd}= -u R + s^{ab} R^T_{ab} + t^{abcd} W_{abcd}.
\end{eqnarray}

Interestingly, when taking the appropriate approximations to study the phenomenological implications of the mgSME, every term containing $t^{abcd}$ vanishes \cite{BaileyKostelecky}, in what has been called the $t$ puzzle \cite{RuiQuentinAlan}. To shed light on this issue, a concrete model in which $t^{abcd}$ is made of $2$-forms, was studied \cite{ABK}, but no physical effects from $t^{abcd}$ were found. To date, the disappearance of the effects associated with $t^{abcd}$ remains an open question. This paper is motivated by such a mysterious disappearance, and its main goal is to determine whether there is a fundamental origin for the $t$ puzzle. As a byproduct of this analysis, some important lessons concerning the gravitational SME sector were learned. In particular, the results reported here may become useful when building the nonminimal gravity SME sector, which includes terms with an arbitrary number of derivatives acting on ${R_{abc}}^d$. In fact, experience from the nonminimal electromagnetic \cite{SME electro nonmin}, neutrino \cite{SME neutrinos nonmin}, and matter \cite{SME fermions nonmin} SME sectors, shows that, to get the nonminimal extensions, it is crucial to identify the physical degrees of freedom in the corresponding minimal sectors.

The equations of motion arising from the variation of the action (\ref{initial action}) with respect to $\phi$, $k^{abcd}$, and $g_{ab}$, respectively, are given by
\begin{eqnarray}
0&=& \frac{\delta S_{\rm matter} }{\delta \phi},\\
0&=& R_{abcd} +2\kappa \frac{\delta S_{\rm coef} }{\delta k^{abcd}},\label{eom k}\\
G^{ab}&=& \frac{1}{2}g^{ab} k^{cdef}R_{cdef} + k^{cde(a}{R_{cde}}^{b)} + 2 \nabla_c \nabla_d k^{c(ab)d}\nonumber\\
&&+ \kappa T_{\rm coef}^{ab}+ \kappa T_{\rm matter}^{ab},\label{metric eom}
\end{eqnarray}
where $G^{ab}=R^{ab}-g^{ab}R/2$ is the Einstein tensor and the energy-momentum tensor associated with $S_{\rm matter}$ is
\begin{equation}
 T_{\rm matter}^{ab} = \frac{1}{2\sqrt{-g}} \frac{\delta S_{\rm matter}}{\delta g_{ab}},
 \end{equation}
which is assumed to have vanishing divergence. The energy-momentum tensor for the coefficients is defined analogously, however, such a tensor typically has a nonzero divergence. In fact, equation (\ref{metric eom}) together with the Bianchi identity yields
\begin{equation}\label{div T coeff}
 \kappa \nabla_{a} T_{\rm coef}^{ab} = - \frac{1}{2}g^{ab}\left(\nabla_a k^{cdef}\right) R_{cdef} - 2 \nabla_a \left(k^{cdea} {R_{cde}}^b \right) ,
\end{equation}
which can also be deduced by requiring $S_{\rm coef}$ to be diffeomorphism invariant. Notice that, to obtain equation (\ref{metric eom}), one has to neglect several boundary terms, which is justified in section \ref{YGH}.

We emphasize that, in contrast to what is done in most mgSME papers, we avoid taking approximations and committing to a concrete $S_{\rm coef}$. Specifically, we do not assume a background geometry, nor we take particular boundary conditions, and we do not separate the SME coefficients into a Lorentz-violating part and its fluctuations. We do neglect quadratic terms in the SME coefficients, which is justified by the empirical fact that Lorentz violation, in all relevant frames, is a small correction to conventional physics.

The paper is organized as follows: Section \ref{Redefinitions} is the core of the manuscript and it describes several analyses where spurious SME-like terms are generated by field redefinitions. In section \ref{Alternatives} alternative fundamental explanations for the $t$ puzzle are described, and in section \ref{concl} the conclusions of this work are given. Finally, Appendix \ref{Notation} describes the notation and conventions, and lists some results used in the paper. 

\section{Field redefinitions}\label{Redefinitions}

It is well established that some SME coefficients arise, from conventional physics, by redefining the dynamical fields \cite{field redef bibliog}. Of course, these spurious coefficients cannot produce new physical effects. The goal of this section is to investigate all possible gravity redefinitions to find the unphysical coefficients in the mgSME. We start from $S_{\rm EH}$ plus a matter and a coefficient action, and we perform all possible redefinitions of the gravitational fields.

\subsection{Metric redefinition}

In this subsection we investigate possible metric redefinitions. For that purpose, we first need to find out how to relate two metric tensors. Let $g_{ab}$ and $\widetilde{g}_{ab}$ be metric tensors, then, there exist two sets of orthonormal basis, $\{e^a_\mu\}$ and $\{\widetilde{e}^a_\mu\}$, such that
\begin{equation}\label{rel tetrads}
 g_{ab}e^a_\mu e^b_\nu = \eta_{\mu\nu} = \widetilde{g}_{ab}\widetilde{e}^a_\mu \widetilde{e}^b_\nu.
\end{equation}
(The notation is explained in the Appendix \ref{Notation}). Let $M_\mu^\nu$ be an invertible spacetime-dependent matrix such that $e^a_\mu=M_\mu^\nu \widetilde{e}^a_\nu$. Then,
\begin{equation}
g^{ab} = M_c^a M_d^b \widetilde{g}^{cd},\label{metric redef}
\end{equation}
where $M_a^b$ is such that, its components, in the $\widetilde{e}^a_\mu$ basis (and its dual) coincide with $M_\mu^\nu$. Observe that $M_\mu^\nu$ has, in principle, $16$ independent components. However, $6$ degrees of freedom correspond to an $SO(1,3)$ rotation that leaves the metric invariant. Thus, we can restrict ourselves to $M_\mu^\nu \in Gl(4)/SO(1,3)$, which has $10$ independent components, as expected.

Let $\widetilde{\nabla}_a $ be the derivative operator associated with $\widetilde{g}_{ab}$, then, using equation (\ref{Riemann transf}) it is possible to see that
\begin{eqnarray}
 \sqrt{-g} R &=&\frac{\sqrt{-\widetilde{g}} }{|M|} M^a_c M^b_d \widetilde{g}^{cd}\left[R_{ab}(\widetilde{\nabla}) -\widetilde{\nabla}_a{C_{eb}}^e \right. \nonumber\\
 && \left.+\widetilde{\nabla}_e{C_{ab}}^e +{C_{ba}}^f{C_{ef}}^e -{C_{be}}^f{C_{af}}^e\right],\label{EH metric redef}
\end{eqnarray}
where $M$ is the determinant of $M_\mu^\nu$. At this point it is tempting to replace ${C_{ab}}^c$ by its expression in terms of $\widetilde{\nabla}_ag_{bc}$, given in equation (\ref{C}), and integrate by parts to get double derivatives. Then, the antisymmetric application of such derivatives can be converted into a Riemann tensor. However, all the double derivatives obtained with this procedure act symmetrically. By inspecting equation (\ref{EH metric redef}) we can conclude that this redefinition allows us to generate the $u$ and $s^{ab}$ terms, but no $t^{abcd}$ term. Recall that $M_\mu^\nu$ has $10$ independent components, which coincides with the number of independent components of $u$ and $s^{ab}$.

As it is well known in scalar-tensor theories \cite{conformal}, a $u$ term can be generated by a conformal transformation, which can be done without appealing to approximations. However, the redefinition involving $s^{ab}$ is more complicated, so we only work to first order in $u$ and $s^{ab}$. At this level of approximation, we can take
\begin{equation}\label{u s redef}
 M_a^b = \left(1+\frac{1}{2}u\right) \delta_a^b +\frac{1}{2}\widetilde{g}_{ac}s^{cb},
\end{equation}
which implies that
\begin{equation}
 |M|^{-1} M^a_c M^b_d \widetilde{g}^{cd} \widetilde{R}_{ab} =(1-u)\widetilde{R} + s^{ab}\widetilde{R}^T_{ab},
\end{equation}
where the geometric tensors with a tilde represent the corresponding tensors associated with $\widetilde{g}_{ab}$. With these expressions, we can see that the Einstein-Hilbert action in terms of $\widetilde{g}_{ab}$ becomes
\begin{equation}\label{redef EH action}
 S_{\rm EH} =\frac{1}{2\kappa} \int d^4x \sqrt{-\widetilde{g}} \left[(1-u)\widetilde{R} + s^{ab}\widetilde{R}^T_{ab} +\widetilde{\nabla}_a T^a\right],
\end{equation}
with 
\begin{equation}
 T^a = -2 \widetilde{g}^{b[a} {C_{bc}}^{c]} = \widetilde{\nabla}_b(4\widetilde{g}^{ab}u-s^{ab}),
\end{equation}
where, to get the second identity, we use equation (\ref{C}). The last term in equation (\ref{redef EH action}) is a total divergence, and, under the appropriate assumptions, it can be ignored. Observe that, even if it is not possible to discard such a term, it can be considered part of $S_{\rm coef}$. After this is done, we can conclude that $u$ and $s^{ab}$ can be generated from the Einstein-Hilbert action by a metric redefinition, which is the main result of this subsection. Also, we have shown that no $t^{abcd}$ term can be produced with a metric redefinition.

\subsection{Implications in other SME sectors}

In this part of the paper we investigate the effects of the metric redefinition on the matter sector. This is achieved by writing a concrete $S_{\rm matter}$ in terms of $\widetilde{g}_{ab}$. We begin by studying the case where $S_{\rm matter}$ contains a gauge field, which, for simplicity, is taken as a free $U(1)$ field $A_a$. In this case
\begin{equation}
 S_{\rm matter}(g,A) =-\frac{1}{4} \int d^4x \sqrt{-g} g^{ac} g^{db }F_{ab}F_{cd},
\end{equation}
where $F_{ab}$ is the standard $U(1)$ field strength
\begin{equation}
 F_{ab} = \nabla_a A_b - \nabla_b A_a . 
\end{equation}
Recall that $F_{ab}$ is independent of the derivative operator, therefore, in terms of $\widetilde{g}_{ab}$ we get
\begin{eqnarray}
 S_{\rm matter}(\widetilde{g},A)& =&-\frac{1}{4} \int d^4x \frac{\sqrt{-\widetilde{g}}}{|M|} M^a_e M^b_f M^c_g M^d_h \widetilde{g}^{eg} \widetilde{g}^{hf}\nonumber\\ &&\times F_{ab}F_{cd}\nonumber\\
 &=& -\frac{1}{4} \int d^4x \sqrt{-\widetilde{g}}\left[\widetilde{g}^{ac} \widetilde{g}^{db}+2s^{ac} \widetilde{g}^{db}\right]\nonumber\\
 && \times F_{ab}F_{cd},
\end{eqnarray}
where the metric redefinition (\ref{u s redef}) is used, and we work to first order in the SME coefficients. Comparing this with the corresponding part of the SME action \cite{Kostelecky04}, we can verify that we have generated part of the term
\begin{equation}
 S_{\rm SME} \supset -\frac{1}{4} \int d^4x \sqrt{-\widetilde{g}} (k_F)^{abcd}F_{ab}F_{cd}.
\end{equation}
In particular, we obtain
\begin{equation}
 (k_F)^{abcd} = s^{a[c} \widetilde{g}^{d]b}-s^{b[c} \widetilde{g}^{d]a}.
\end{equation}

We turn to consider the effects of the metric redefinition on matter fields. The action for a free Dirac fermion $\psi$ in a curved background is given by
\begin{eqnarray}
 S_{\rm matter}(g,\psi) &=&\int d^4x e \left[\frac{i}{2} e^a_\mu \bar{\psi} \gamma^\mu \overleftrightarrow{\nabla}_a \psi -m \bar{\psi}\psi\right]\nonumber\\
 &=& \int d^4x e \left[\frac{i}{2} e^a_\mu \bar{\psi} \gamma^\mu \overleftrightarrow{\partial}_a \psi -m \bar{\psi}\psi\right.\nonumber\\
 && \left. - \frac{1}{4}e^a_\mu \omega_{a\rho\sigma}\eta_{\alpha\beta}\epsilon^{\mu\rho\sigma\alpha} \bar{\psi}\gamma_5 \gamma^\beta \psi\right],\label{action standard fermion 2}
\end{eqnarray}
where the notation is explained in Appendix \ref{Notation}. Following the strategy outlined above, we want to write this action in terms of $\widetilde{g}_{ab}$, and the tetrad $\widetilde{e}^a_\mu$ associated with it. Let $\widetilde{\omega}_{a\mu\nu} = \widetilde{g}_{bc}\widetilde{e}^b_\mu \widetilde{\nabla}_a \widetilde{e}^c_\nu$, then 
\begin{eqnarray}
 e^a_\mu\omega_{a\rho\sigma} &=&\widetilde{e}^b_\mu M_b^a \widetilde{\omega}_{a\rho\sigma}+ \widetilde{g}_{cd} \widetilde{e}^b_\mu \widetilde{e}^c_\rho \widetilde{e}^e_\sigma M_b^a(M^{-1})_f^d \nonumber\\
 && \times \left(\widetilde{\nabla}_a M_e^f+ M_e^g {C_{ag}}^f\right).
 \end{eqnarray}
Note that, to first order in the SME coefficients,
 \begin{eqnarray}
\widetilde{g}_{cd} \widetilde{e}^b_\mu \widetilde{e}^c_\rho \widetilde{e}^e_\sigma M_b^a(M^{-1})_f^d \left(\widetilde{\nabla}_a M_e^f+ M_e^g {C_{ag}}^f\right)
 \end{eqnarray}
is symmetric under the interchange of $\mu$ and $\sigma$, and thus, it vanishes when contracted with $\epsilon^{\mu\rho\sigma\alpha}$, therefore,
\begin{equation}
 S_{\rm matter}(\widetilde{g},\psi)= \int d^4x \frac{\widetilde{e}}{|M|} \left[\frac{i}{2} \widetilde{e}^b_\mu M_b^a \bar{\psi} \gamma^\mu \overleftrightarrow{\widetilde{\nabla}}_a \psi- m \bar{\psi}\psi\right],\label{action psi redef}
\end{equation}
where $\widetilde{\nabla}_a$ is defined by equations (\ref{nabla psi1}-\ref{nabla psi2}) but with $\widetilde{\omega}_{a\mu\nu}$ replacing $\omega_{a\mu\nu}$. The action (\ref{action psi redef}) has the unpleasant feature of describing an SME Dirac fermion with a spacetime-dependent mass. However, rescaling the fermionic field by $\psi = \sqrt{|M|}\chi$ yields
\begin{equation}
 S_{\rm matter}(\widetilde{g},\chi)= \int d^4x \widetilde{e} \left[\frac{i}{2} \widetilde{e}^b_\mu M_b^a \bar{\chi} \gamma^\mu \overleftrightarrow{\widetilde{\nabla}}_a \chi- m \bar{\chi}\chi\right],\label{action chi redef}
\end{equation}
This expression must be compared with the term in the SME action containing the $c_{ab}$ coefficient, which, in terms of the tetrad $\widetilde{e}^a_\mu$ and the Dirac fermion $\chi$, reads \cite{Kostelecky04}
\begin{equation}
 S_{\rm SME} \supset \int d^4x \widetilde{e}\left(\frac{-i}{2}\right)\widetilde{e}^b_\mu \widetilde{g}^{ac} c_{bc} \bar{\chi}\gamma^\mu \overleftrightarrow{\widetilde{\nabla}}_a \chi.
\end{equation}
By direct inspection we can verify that the $c_{ab}$ coefficient generated by the metric redefinition is given by
\begin{equation}
c_{ab} = -\widetilde{g}_{bc} M_a^c=- \frac{1}{2} \left(\widetilde{g}_{ab}u + \widetilde{g}_{ac}\widetilde{g}_{db}s^{cd}\right).
\end{equation}

As a last example we consider the QED interaction term
\begin{equation}\label{QED interaction action}
 S_{\rm matter}(g,A,\psi)=\int d^4 x e e^a_\mu A_a \bar{\psi}\gamma^\mu \psi.
\end{equation}
Under the metric redefinition we get
\begin{equation}
S_{\rm matter}(\widetilde{g},A,\psi)= \int d^4 x \frac{\widetilde{e}}{|M|} M_b^a \widetilde{e}^b_\mu A_a \bar{\psi}\gamma^\mu \psi .
\end{equation}
Moreover, we can rescale the fermionic field as before, obtaining
\begin{equation}
S_{\rm matter}(\widetilde{g},A,\chi)= \int d^4 x \widetilde{e} M_b^a \widetilde{e}^b_\mu A_a \bar{\chi}\gamma^\mu \chi .
\end{equation}
which allows us to verify that we recover the standard $U(1)$ gauge invariance. Observe that the term generated in this case must also be part of the SME, however, the interaction sectors of the SME have not been classified.

From the analysis presented in this subsection we can conclude that the $u$ and $s^{ab}$ coefficients generated by a metric redefinition also affect the matter sectors. Obvious modifications to this analysis can be used to cancel coefficients in one sector in favor of coefficients in other sectors. In this sense it can be said that the mgSME coefficients $u$ and $s^{ab}$ can be moved into the matter sectors. Note that, even in the absence of matter fields, $u$ and $s^{ab}$ are not completely unphysical since they appear in the action term $S_{\rm coef}$; this was originally noticed in the context of Einstein-{\AE}ther theory \cite{Aether}. We want to remark that, if the linearized gravity limit is taken, the expressions provided here match the previously known results of Ref.~\onlinecite{KosteleckyTasson}.

So far we have analyzed metric redefinitions. However, this study offers no solution to the $t$ puzzle. To make sure the $t$ term cannot be generated with a field redefinition, we must study the most general type of redefinitions. In particular, in the next subsection we analyze independent redefinitions of the metric and the derivative operator.

\subsection{Palatini redefinitions}

A remarkable feature of general relativity is that it is possible to take the metric and the derivative operator as \textit{a priori} independent variables, in what is known as the Palatini approach \cite{Palatini}. Inspired by this result, we investigate the viability of performing independent redefinitions to the metric and the derivative operator in the mgSME. Since this redefinition is more general than the metric redefinition, it has the potential of generating a $t^{abcd}$ term. However, before doing this generalized redefinition, we need to prove that the mgSME action can be treated \textit{\`a la} Palatini. This is what we do next.

Let $\hat{\nabla}_a$ be a derivative operator that is independent of $g_{ab}$. Then, the mgSME action in a Palatini formalism is
\begin{eqnarray}
S_{\rm mgSME}&=& \frac{1}{2\kappa}\int d^4 x \sqrt{-g} (g^{ca}\delta^b_d + g_{de} k^{abce}){R_{abc}}^d(\hat{\nabla})\nonumber\\
&& + S_{\rm coef}+ S_{\rm matter}.
\end{eqnarray}
The equation of motion obtained from a metric variation is
\begin{eqnarray}
0&=&\left(\frac{1}{2} g^{ab}g^{cd} - g^{a(c}g^{d)b}\right) R_{cd}(\hat{\nabla})\nonumber\\
&&+ \frac{1}{2} g^{ab}g_{fg} k^{cdef}{R_{cde}}^g(\hat{\nabla})\nonumber\\
&& + k^{cde(a}{R_{cde}}^{b)}(\hat{\nabla})+ \kappa T_{\rm coef}^{ab}+ \kappa T_{\rm matter}^{ab},\label{eom Palatini SME}
\end{eqnarray}
where emphasis is made on the fact that $R_{ab}(\hat{\nabla})$ need not be symmetric. To get the equation of motion associated with a variation of the derivative operator we define ${\hat{C}_{ab}}^c$ as the connection linking $\nabla_a$ and $\hat{\nabla}_a$ and we take ${\hat{C}_{ab}}^c$ to be the independent physical field. Using an equation similar to equation (\ref{Riemann transf}) for $\hat{\nabla}_a$, and assuming, as is customary, that neither $S_{\rm coef}$ nor $S_{\rm matter}$ depend on the derivative operator, we get
\begin{eqnarray}
0&=& g^{de} {\hat{C}_{de}}^{(a} \delta^{b)}_c - 2g^{d(a}{\hat{C}_{cd}}^{b)} + g^{ab} {\hat{C}_{cd}}^d + 2g_{cd}\nabla_e k^{d(ab)e}\nonumber\\
&&- 2 g_{ef} k^{d(ab)f}{\hat{C}_{cd}}^e + 2g_{cf} k^{dfe(a}{\hat{C}_{de}}^{b)} .
\end{eqnarray}
To obtain an equation that can be compared with equation (\ref{metric eom}) we need to solve ${\hat{C}_{ab}}^c$ as a function of $g_{ab}$ and $k^{abcd}$, and substitute it into equation (\ref{eom Palatini SME}). It is possible to see that, to first order in the SME coefficients, the connection satisfies
\begin{equation}
 {\hat{C}_{ab}}^c = \left(g_{ab}g_{ef} \delta_g^c-2g_{e(a}g_{b)f}\delta_g^c - \frac{4}{3}g_{ef}g_{g(a}\delta_{b)}^c \right)\nabla_d k^{defg}.
\end{equation}
Inserting this result into equation (\ref{eom Palatini SME}) allows us to verify that it coincides with equation (\ref{metric eom}). In addition, the equations of motion arising from $S_{\rm coef}$ and $S_{\rm matter}$ also coincide since, by assumption, such actions are independent from the derivative operator. Therefore, we can conclude that, to first order in the SME coefficients, the Palatini variation of the action (\ref{initial action}) produces equivalent equations of motion than the standard approach.

We can now proceed to investigate the independent redefinitions of the metric and the derivative operator. Let us take the Palatini-Einstein-Hilbert action
\begin{equation}
S_{\rm PEH}= \frac{1}{2\kappa}\int d^4 x \sqrt{-g} g^{ab}R_{ab}(\hat{\nabla})
\end{equation}
as our starting point. Under a metric redefinition that leaves the derivative operator unchanged, we get
\begin{equation}
 \sqrt{-g}g^{ab}R_{ab}(\hat{\nabla}) = \frac{\sqrt{-\widetilde{g}}}{|M|} M^a_c M^b_d\widetilde{g}^{cd} R_{ab}(\hat{\nabla}).
\end{equation}
Clearly, it is possible to choose $M_a^b$ as in equation (\ref{u s redef}) to generate the $u$ and $s^{ab}$ terms. On the other hand, changing the derivative operator from $\hat{\nabla}_a$ to $\breve{\nabla}_a$, without affecting the metric, yields
\begin{eqnarray}
 \sqrt{-g}g^{ab}R_{ab}(\hat{\nabla}) &=& \sqrt{-g}g^{ab}\left[R_{ab}(\breve{\nabla}) - 2 \breve{\nabla}_{[a}{\breve{C}_{c]b}}^c\right. \nonumber\\
 && \left.+2 {\breve{C}_{b[a}}^d{\breve{C}_{c]d}}^c\right],
\end{eqnarray}
where ${\breve{C}_{ab}}^c$ is the connection relating the derivative operators under consideration. Upon inspection, it can be concluded that no SME terms are generated under this redefinition. In fact, by converting the second term into a total divergence, this expression can be thought of as a standard Einstein-Hilbert-Palatini action plus terms quadratic in the connections that may be absorbed into $S_{\rm coef}$. One could naively consider a redefinition where ${\breve{C}_{ab}}^c = \breve{\nabla}_{(a} l_{b)}^c$, for a generic tensor $l_a^b$, to obtain terms with two antisymmetric derivatives which, in turn, can be written as a curvature tensor. However, such a redefinition is not invertible, and thus, it does not leave the physics invariant. Finally, one can consider simultaneous and independent redefinitions of the metric and the derivative operator. This is the most general redefinition of the gravitational degrees of freedom. Nevertheless, it is not particularly illuminating for our purposes. In fact, such a redefinition gives
\begin{eqnarray}
 \sqrt{-g}g^{ab}R_{ab}(\hat{\nabla}) &=& \frac{\sqrt{-\widetilde{g}}}{|M|} M^a_c M^b_d\widetilde{g}^{cd}\left[R_{ab}(\breve{\nabla}) \right. \nonumber\\
 && \left.- 2 \breve{\nabla}_{[a}{\breve{C}_{c]b}}^c+2 {\breve{C}_{b[a}}^d{\breve{C}_{c]d}}^c\right],
\end{eqnarray}
where no $t^{abcd}$ term is produced. 

In principle, it is also possible to study Palatini-like redefinitions in the language of tetrads by considering the spin connection $\omega_{a\mu\nu}$ as an independent field from $e^a_\mu$. However, it is not clear how to redefine the spin connection and, by analyzing some concrete redefinitions, it is possible to see that there is no clear advantage with respect to the standard Palatini approach. An interesting line of research that could be related to the $t$ puzzle is to study general canonical transformations, which calls for the mgSME Hamiltonian. This is technically involved, even in general relativity, and it is thus left for future work.

\section{Alternative explanations}\label{Alternatives}

\subsection{Boundary terms}\label{YGH}

It is well known that, when spacetime $M$ has a boundary $\partial M$, the Einstein-Hilbert action needs to be corrected to produce the Einstein field equations by varying the action, while keeping the derivatives of the metric variation at $\partial M$ arbitrary. In general relativity, such a correction is known as the York-Gibbons-Hawking term \cite{HG term}. Moreover, in the phenomenological applications of the mgSME, spacetime is conformally flat, which implies that it has a boundary. Thus, it is important to see if the mgSME action can be corrected by a corresponding boundary term. This is what we do in this subsection. For simplicity, we only consider spacetimes with non-null boundaries.

Consider the following piece of the mgSME action variation:
\begin{eqnarray}
 \delta S_{\rm mgSME} &\supset& \frac{1}{2\kappa}\int_M d^4x \sqrt{-g} (g^{ca}\delta_d^{b}+{k^{abc}}_d )\delta {R_{abc}}^d \nonumber\\
 &=&\frac{1}{\kappa}\int_M d^4x \sqrt{-g} (\nabla_c\nabla_d k^{cabd} )\delta g_{ab}\nonumber\\
 &&+\frac{1}{\kappa}\int_{\partial M} d^3x \sqrt{|h|}n_c (g^{a[c}g^{b]d}+k^{cabd} )\nabla_d \delta g_{ab}\nonumber\\
 &&-\frac{1}{\kappa}\int_{\partial M} d^3x \sqrt{|h|} n_c (\nabla_d k^{cabd} )\delta g_{ab},\label{delta S w boundary}
 \end{eqnarray}
where $n_a$ is the unit vector normal to $\partial M$,
\begin{equation}
h_{ab} = g_{ab} \pm n_a n_b 
\end{equation}
is the induced metric at $\partial M$ (the sign depends on whether the boundary is spacelike or timelike), and $h$ is the determinant of the components of $h_{ab}$. We recognize the first term in equation (\ref{delta S w boundary}) as one of the contributions to the equations of motion (\ref{metric eom}), and we note that the last term is zero since the metric variation at $\partial M$ vanishes by assumption. However, the second term in equation (\ref{delta S w boundary}) does not vanish.

Let $K_{ab}=h_a^c \nabla_c n_b$ be the extrinsic curvature of the boundary, then
\begin{equation}
\delta K_{ab} =-h_a^c n_d \delta {C_{cb}}^d =\frac{1}{2}h_a^c n^d\nabla_d \delta g_{bc},
\end{equation}
where the variation is done fixing $\delta g_{ab}$ at $\partial M$. This result allows us to write the integrand of the second term as
\begin{eqnarray}
&&n_c (g^{a[c}g^{b]d}+k^{cabd} )\nabla_d \delta g_{ab}\nonumber\\
&=& -\frac{1}{2}(h^{ab} \pm 2 k^{cabd}n_c n_d ) n^e\nabla_e \delta g_{ab}\nonumber\\
&=&-\left(h^{ab} \pm 2 n_c n_d k^{cabd} \right)\delta K_{ab}\nonumber\\
&=&-\delta[\left(h^{ab} \pm 2 n_c n_d k^{cabd} \right) K_{ab}],\label{boundary term}
\end{eqnarray}
where we use that the tangential derivatives of $\delta g_{ab}$ vanish. Thus, to cancel the term (\ref{boundary term}), it is necessary to add to the action (\ref{initial action}) the boundary piece
\begin{equation}
 \Delta S_{\rm mgSME} = \frac{1}{\kappa}\int_{\partial M} d^3x \sqrt{|h|}\left(K \pm 2 n_a n_d k^{abcd} K_{bc} \right). 
\end{equation}
Observe that the first term in $\Delta S_{\rm mgSME}$ is the York-Gibbons-Hawking term, while the second depends on the SME coefficient. Note that the equations of motion associated with the variation of the SME coefficients do not change by the presence of $\Delta S_{\rm mgSME}$ since those equations are obtained by varying the action with the coefficient fixed at the boundary.

The remarkable conclusion is that, in the cases where there are no additional metric-variation derivatives coming from $S_{\rm coef}$, the mgSME action can be corrected to cancel the contribution of the boundary terms, which is not a generic feature in geometric theories. For example, this analysis can be repeated for actions that are nonlinear in $R_{abcd}$, like some terms in the nonminimal SME sector, or an $f(R)$ theory \cite{f(R)}, and it is not hard to see that in such cases it is impossible to correct the action with a boundary term. In fact, the same conclusion can be reached when the $f(R)$ is translated to the language of a nonminimally coupled scalar field \cite{Luisa}.

\subsection{Lanczos-like tensor}

It has been shown \cite{Bampi} that, in $4$ spacetime dimensions, any analytic tensor with the index symmetries of the Weyl tensor can be written in terms of derivatives of a tensor $H^{abc}$, which, in the particular case of the Weyl tensor, is called the Lanczos tensor \cite{Lanczos}. In particular, the $t^{abcd}$ coefficient can be written as
\begin{eqnarray}
 t^{abcd} &=& g^{de}\nabla_e H^{abc} - g^{ce}\nabla_e H^{abd} + g^{be}\nabla_e H^{cda}\nonumber\\
 && - g^{ae}\nabla_e H^{cdb}+g^{ac}\nabla_e H^{ebd}-g^{ad}\nabla_e H^{ebc}\nonumber\\
&&+g^{bd}\nabla_e H^{eac}-g^{bc}\nabla_e H^{ead}.\label{Lanczos2}
\end{eqnarray}
where $H^{abc} =- H^{bac}$, $H^{[abc]} =0 $, $g_{bc}H^{abc}=0 $, and $\nabla_c H^{abc}=0$. Notice that $H^{abc}$ has $10$ independent components, coinciding with the number of independent components of any tensor sharing the index symmetries of the Weyl tensor.

If we replace $t^{abcd}$, in the mgSME action, by equation (\ref{Lanczos2}), we obtain
\begin{eqnarray}
 S_{\rm LV}&=& \frac{1}{2\kappa}\int d^4 x \sqrt{-g} \left[ -u R + s^{ab} R^T_{ab} -4 H^{abc}\nabla_d{W_{abc}}^d\right]\nonumber\\
 &=& \frac{1}{2\kappa}\int d^4 x \sqrt{-g} \left[ -u R + s^{ab} R^T_{ab} +4 H^{abc}\nabla_aR_{bc}\right].\nonumber\\\label{LV action with H}
\end{eqnarray}
Here, we have integrated by parts neglecting the surface term, and used the Bianchi identity in the form
\begin{equation}
 \nabla_d{W_{abc}}^d = - \nabla_{[a}R_{b]c} - \frac{1}{6}g_{c[a}\nabla_{b]}R.
\end{equation}
The important point is that we have shown that an analytic $t^{abcd}$ coefficient can always be mapped into a dimension-$5$ operator $H^{abc}$ acting on $\nabla_aR_{bc}$. As noticed in Ref.~\onlinecite{RuiQuentinAlan}, such operators, in the nonrelativistic weak-gravity approximation, lead to pseudovector contributions in the gravitational force, which, in turn, generate nonphysical self-accelerations. Thus, in realistic models, dimension-$5$ operators have to be removed. Observe that a similar mechanism that relates coefficients of different dimensions has been discovered in the nonminimal SME fermion sector \cite{SME fermions nonmin}.

It is tempting to perform an additional integration by parts in equation (\ref{LV action with H}) to get an effective coefficient $s_{\rm eff}^{ab} = s^{ab}- 4 \nabla_c H^{cab}$ contracted with $R^T_{ab}$. However, such an effective coefficient depends, through the covariant derivative, on the metric. In addition, $s_{\rm eff}^{ab}$ cannot be redefined away since the necessary redefinition would be noninvertible.

\subsection{Other ideas}

\subsubsection{Initial value formulation}

The equation of motion (\ref{metric eom}) includes, at most, two time derivatives of the metric. This is a desirable condition as it implies that the metric evolution only depends on its initial value and its initial time derivative. However, those are not the only conditions that a theory should satisfy to have a proper evolution. In fact, any viable physical theory should have a well-posed Cauchy problem, namely, given suitable initial data, which may be subject to constraints, the evolution should be unique, causal, and continuous (for a formal description of such conditions see Ref.~\onlinecite{Wald}). Notice that causality has been studied in the SME, but only in a flat spacetime \cite{Ralf}. The goal of this part of the paper is not to study the Cauchy problem of the mgSME, but to pinpoint potential difficulties of such a research program, and to argue that such an analysis could reveal obstructions related to the $t$ puzzle. 

In general relativity, not all the components of the Einstein equations are evolution equations; some are dynamical constraints. Analogously, in the mgSME there are several constraints, however, such relations are not simply some components of the equations of motion. Therefore, to identify the constraints, it is necessary to perform the Dirac algorithm \cite{Dirac}, which, in turn, calls for the Hamiltonian formulation of the mgSME. As part of this analysis, one would need to verify that all the pieces of the action (\ref{initial action}), including $S_{\rm coef}$ and $S_{\rm matter}$, have a well-posed Cauchy problem. Remarkably, most natural actions for tensors of spin larger than $1$ have an ill-posed Cauchy problem \cite{Wald}, and $t^{abcd}$ cannot be built out of tensors of spin smaller or equal to $1$. Therefore, this analysis will certainly impose restrictions on the dynamics of $t^{abcd}$.

\subsubsection{Nondual basis}

The tensor operator of index contraction maps a $(k,l)$ tensor to a $(k-1,l-1)$ tensor by selecting an upper and lower tensor entry, inserting in them, one at a time, the components of a basis and its corresponding dual, and adding the resulting tensors. The idea reported in this part of the paper is to analyze if an SME-like term arises from the Einstein-Hilbert action if the contractions are done with a nondual basis. It turns out that such a scheme generates a $k^{abcd}R_{abcd}$ term, however, what plays the role of $k^{abcd}$ are some combinations of the inverse-metric components, for which no mechanism to brake local Lorentz invariance spontaneously is given. Thus, it is not possible to generate spurious SME coefficients by using a nondual basis.

\subsubsection{Naive perturbative treatment}

Here we present a perturbative treatment of the mgSME to show that such perturbations can impose consistency conditions in lower-order equations. Let us take an $\epsilon$ out of $k^{abcd}$ and write the metric as $g_{ab} =g_{ab}^{(0)} + \epsilon g_{ab}^{(1)}$, where the zeroth order metric is a solution of Einstein equation with the matter fields acting as the source. Furthermore, assume that in equation (\ref{eom k}), the variation of $S_{\rm coef}$ is of order $\epsilon$. Thus, equation (\ref{eom k}), at zeroth order in $\epsilon$, yields $R_{abcd}(g_{ab}^{(0)}) =0$, forcing the matter fields to behave as test fields, in contradiction to the previous assumptions.

If the irreducible components of $k^{abcd}$ are separately considered, the restriction discussed above relaxes significantly. For example, if only $s^{ab}$ is considered, then spacetime has to be Ricci flat, which allows the background geometry to be a vacuum solution of Einstein equations. On the other hand, if only $t^{abcd}$ is considered, then the background geometry is forced to be conformally flat. This analysis is only presented to show that, in the mgSME, perturbative treatments tend to generate inconsistencies, which should be studied in each particular case.

\section{Conclusions}\label{concl}

We investigated possible fundamental reasons behind the $t$ puzzle, none of which actually solves the problem. However, during this process, several interesting conclusions were reached, and some gaps in the understanding of the mgSME were closed.

Field redefinitions have been used in several SME sectors to pinpoint the coefficients that have physical implications. Therefore, it was natural to undertake the corresponding study in the gravitational sector. It was demonstrated that the $u$ and $s^{ab}$ coefficients can indeed be moved into other SME sectors through field redefinitions. This was done without appealing to the linearized gravity approximation, or other restricting assumptions, except to work to linear order in the SME coefficients. In addition, it was proven that the mgSME can be treated in a Palatini formalism. This, by itself, is an interesting result, which allowed us to analyze field redefinitions where the metric and the derivative operator are independently redefined. However, even with these more general redefinitions, it is not possible to produce a $t^{abcd}$ term.

Since we proved that field redefinitions cannot solve the $t$ puzzle, we investigated other explanations. First, it was shown that boundary contributions that arise when varying the mgSME action can be canceled with a generalization of the well-known York-Gibbons-Hawking term. Also, rewriting the $t^{abcd}$ coefficient in terms of the derivative of a Lanczos-like tensor helped us show that, at the phenomenological level, such a coefficient could generate unphysical self-accelerations, and that the $t^{abcd}$ term could be mapped onto a metric-dependent $s^{ab}$ coefficient. On the other hand, the relevance of studying the Cauchy problem for the mgSME is emphasized, together with some potential problems associated with $t^{abcd}$. Finally, other ideas are explored, like the possibility that $t^{abcd}$ appears through a nondual basis, and that restrictions could arise when analyzing the mgSME perturbatively.

The results from this paper could become relevant when trying to construct the nonminimal gravity SME sector. For example, the Palatini vs standard analysis suggests that the equivalence of those approaches could not hold in the nonminimal extension. Furthermore, it is mentioned that, for some nonminimal terms, it is impossible to modify the action to cancel the boundary-term effects. It should be stressed that much of the fundamental studies that are left out of this paper call for the Hamiltonian formulation of the mgSME. Such a formulation could allow us to study generic canonical transformations, which can be thought of as more general field redefinitions, and thus have the chance of generating a $t^{abcd}$ term. On the other hand, the dynamical constraints of the mgSME, which are needed to analyze the Cauchy problem, are automatically obtained in the Hamiltonian formalism.

Finally, we want to stress that the results of this paper suggest that the $t$ puzzle has no fundamental origin. The alternative is that the puzzle is an artifact of the phenomenological approximations. Hence, it would be interesting to search for the effects of $t^{abcd}$ in other phenomenological situations and using different approximations.

\begin{acknowledgments}
I thank Alan Kosteleck\'y for pointing the $t$ puzzle out to me and for many discussions on the subject. Also, I benefited from interacting with A. Corichi, E. Montoya, M. Salgado, and D. Sudarsky. This work was supported in part by U.S. DOE Grant No. {DE}-SC0010120 and by the Indiana University Center for Spacetime Symmetries.
\end{acknowledgments} 

\appendix

\section{Notation and conventions}\label{Notation}

We use the notation and conventions of Ref.~\onlinecite{Wald}. Specifically, we work in $4$ spacetime dimensions, we use a metric of signature $-2$, and we use Latin letters to denote abstract indexes. Also, the derivative operator $\nabla_a$ is associated with $g_{ab}$, namely, $\nabla_a g_{bc}=0$. Moreover, since several metrics are simultaneously used, no indexes are raised or lowered by a metric.

Any spacetime admits an infinite collection of derivative operators. It can be shown that the difference of any two derivative operators, $\nabla_a$ and $\widetilde{\nabla}_a$, when acting on a generic covector $\omega_a$, satisfies
\begin{equation}
 (\nabla_a - \widetilde{\nabla}_a)\omega_b =- {C_{ab}}^c(\nabla,\widetilde{\nabla})\omega_c,\label{def C}
\end{equation}
where ${C_{ab}}^c$ is a tensor which is symmetric in its lower indexes. This condition can be generalized as
\begin{equation}
 (\nabla_a - \widetilde{\nabla}_a)T^{a_1\ldots}_{b_1\ldots} = -{C_{ab_1}}^cT^{a_1\ldots}_{c\ldots}-\cdots+{C_{ac}}^{a_1}T^{c\ldots}_{b_1\ldots}+\cdots.
\end{equation}

The Riemann tensor is defined as the result of acting twice with a derivative operator in an antisymmetric way. Using the expression (\ref{def C}), it is possible to obtain
\begin{equation}\label{Riemann transf}
 {R_{abc}}^d(\nabla) = {R_{abc}}^d(\widetilde{\nabla}) -2\widetilde{\nabla}_{[a}{C_{b]c}}^d+2{C_{c[a}}^e{C_{b]e}}^d.
\end{equation}
Note that, when $\widetilde{\nabla}_a$ is the derivative operator associated with a coordinate system, namely, a partial derivative, ${C_{ab}}^c$ becomes the Christoffel symbols and we get the standard expression for the Riemann tensor. In addition, following the standard derivation of the Christoffel symbols, it is possible to show that
\begin{equation}\label{C}
{C_{ab}}^c = \frac{1}{2}g^{cd}(\widetilde{\nabla}_a g_{bd}+\widetilde{\nabla}_b g_{ad}-\widetilde{\nabla}_d g_{ab}). 
\end{equation}

The spacetime geometry can be encoded by an orthonormal basis $e^a_\mu$, namely, a basis such that $g_{ab}e^a_\mu e^b_\nu = \eta_{\mu\nu}$, where Greek indexes should be understood as labels for the different elements of the basis, and $\eta_{\mu\nu}$ is the matrix whose nonzero elements are ${\rm diag}(-1,1,1,1)$. The spin connection encodes how tetrads in neighboring tangent spaces are compared and are given by $\omega_{a\mu\nu} = g_{bc}e^b_\mu \nabla_a e^c_\nu$. By the way, observe that the derivative operators used in this paper differ from the derivative used in most SME papers, like Ref.~\onlinecite{Kostelecky04}, since their derivative, denoted by $D_a$, annihilates the tetrads. It is also possible to show that the determinant of the dual tetrad components, $e$, is equivalent to $\sqrt{-g}$.

The Dirac matrices $\gamma^\mu$ follow the convention that $\{\gamma^\mu,\gamma^\nu\}=-2 \eta^{\mu\nu}$. The covariant derivative acting on a Dirac spinor gives
\begin{eqnarray}
\nabla_a \psi = \partial_a \psi + \frac{i}{4} \omega_{a\mu\nu}\sigma^{\mu\nu}\psi\label{nabla psi1},\\
\nabla_a \bar{\psi} = \partial_a \bar{\psi}- \frac{i}{4} \omega_{a\mu\nu}\bar{\psi}\sigma^{\mu\nu}\label{nabla psi2},
\end{eqnarray}
with $\sigma^{\mu\nu} = i[\gamma^\mu,\gamma^\nu]/2$ and $\bar{\psi} = \psi^\dagger \gamma^0$. For any derivative operator, it is customary to use the following notation:
\begin{equation}
\bar{A} \overleftrightarrow{\nabla}_a B = \bar{A} \nabla_a B - (\nabla_a \bar{A}) B.
\end{equation}
Finally, we introduce $\epsilon^{\mu\nu\rho\sigma}$ as the totally antisymmetric symbol such that $\epsilon^{0123} = -1$ and $\gamma_5 = i \gamma^0\gamma^1\gamma^2\gamma^3$. With these definitions it is possible to prove the following identity:
\begin{equation}
\{ \gamma^\mu, \sigma^{\rho\sigma} \} = 2 \eta_{\alpha\beta} \epsilon^{\mu\rho\sigma\alpha}\gamma_5 \gamma^\beta.
\end{equation}

\end{document}